\documentclass[12pt,preprint]{aastex}

\newcommand\beq{\begin{equation}} \newcommand\eeq{\end{equation}}

\def\mpc{Mpc}
\def\a168{Abell~168}
\def\aap{A\&A} 

\def\etal{et al.\ }
\def\ie{i.e., }
\def\eg{e.g., }
\def\kms{$\rm{km\ s}^{-1}$}
\def\Msun{$M_{\odot}$}
\def\arcmin{$^{\prime}$ }

\received{-- ----} \accepted{-- ----}

\shorttitle{} \shortauthors{}

\begin{document}

{ \title{Large-Field Multicolor Study of \a168 : \\
Subclusters, Dynamics and Luminosity Functions}

\author{ Yanbin Yang\altaffilmark{1}, Xu Zhou\altaffilmark{1},
Qirong Yuan\altaffilmark{1,2}, Zhaoji Jiang\altaffilmark{1},
Jun Ma\altaffilmark{1}, Hong Wu\altaffilmark{1},
Jiansheng Chen\altaffilmark{1} }

\altaffiltext{1}{National Astronomical Observatories, Chinese Academy
of Sciences, Beijing, 100012, P.\ R.\ China }
\altaffiltext{2}{Department of Physics, Nanjing Normal University,
                 NingHai Road 122, Nanjing, 210097, P.\ R.\ China}
\email{yyb@bac.pku.edu.cn} 
\email{zhouxu@bac.pku.edu.cn}

\begin{abstract}
This paper presents a multicolor study of the nearby cluster of
galaxies \a168 ($z=0.045$) with 13 intermediate-band filters in the
Beijing-Arizona-Taiwan-Connecticut (BATC) filter system. After a
cross-identification between the photometric data obtained from the
BATC and the Sloan Digital Sky Survey (SDSS), a catalog containing 1553
galaxies down to $r'<20.0$ mag is achieved, which includes  121
spectroscopically confirmed member galaxies. The technique of
photometric redshift has been applied to all these galaxies with
combined 18-band (13 from BATC and  5 from SDSS) spectral energy
distributions (SEDs), in order to perform a faint membership selection
in \a168.  As a result, 255 galaxies are newly selected as the member
candidates.

Based on the enlarged sample of cluster galaxies, the spatial
distribution and dynamics of Abell 168 are investigated.   In the
light of  the spatial distribution of the member galaxies and the
$0.2-3.5$~keV  X-ray image by Einstein observatory,  it seems that
\a168  consists of two merging subclusters  with a relative radial velocity 
of $264\pm 142$~\kms. 
With the help of ROSTAT software, a detailed investigation of 
the dynamics shows the intrinsic difference in the velocity distributions
for these two subclusters. Under a linear two-body model, 
they are found to be a gravitationally bound system
with 92\% probability. 
The slight deviation of
the local velocity distribution from the overall distribution in the
central region may suggest a picture of head-on collision in the
projection plane.  By using a {\sl double} Schechter function,  the
close examination of the luminosity functions reveals the different
galaxy content of these two subclusters,  which implies the merging
process is possibly at the early stage.

\end{abstract}


\section{INTRODUCTION}

\label{secintr}
The increasing volume of observational data for galaxy clusters (GCs)
in multiple wavebands provides important constraints upon the
formation of large scale structures and cosmology (Bahcall 1988;
Dressler \& Gunn 1988; Schindler 2001). The investigations on the
luminosity function and the spectral features of member galaxies in a
rich cluster will help us to understand not only the structures and
dynamics of the whole galaxy cluster but also the formation and
evolution of galaxies under such a dense environment (Dressler 1984;
Poggianti 2002).

Spectroscopic observation in the optical band is still the most
straightforward and powerful approach to the membership
determination. However, the reliable spectroscopies for the faint
galaxies in a cluster is still an unattainable goal for the large
telescopes.  With the development of photometric redshift technique,
the redshifts of faint galaxies can be significantly determined
according to their SEDs.  The BATC sky survey is specially designed
for this purpose.  In this paper we present a large-field multicolor
study of the galaxy cluster \a168 based on the BATC observations with
13 intermediate-band filters, which covers the whole optical band.
Improvement of the accuracy in photometric redshift estimate with our
multicolor system allows us to study the spatial and the dynamic
properties of the nearby clusters of galaxies (Yuan \etal 2001; Yuan
\etal 2003).

\a168 is a nearby ($z=0.045$) cluster of galaxies located at ${01^{\rm
h}15^{\rm m}10^{\rm s}.2}$~+$00^\circ12^{\prime}18^{\prime\prime}$
(J2000.0) with richness of I$\!$I$-$I$\!$I$\!$I (BM classification).
Its distance module is 36.45 by assuming $q_0=0.5$ and $H_0=75\rm\,
km\,s^{-1}Mpc^{-1}$ (also used throughout this paper).  The luminosity
function of \a168 was remarked  as the best example of non-Schechter
form by Dressler (1978), which was confirmed by Oegerle, Ernst \&
Hoessel (1986). Based on the positional discrepancy of $\sim 10'$
(\ie $\sim$ 540 kpc) between the X-ray and the optical center, 
Ulmer, Wirth \& Kowalski (1992) suggested that \a168 was
formed by the collision of two clusters with approximately equivalent
sizes.  It is reported that the X-ray emission of \a168 has two peaks
with surface brightness of 9$\times10^{-14}$ and
3$\times10^{-14}{\rm\,erg\, s^{-1}cm^{-2}}$, respectively (Tomita \etal 1996). 
They supposed that  the star-forming processes will be
triggered when gas-rich galaxies rush into the higher density
regions. However,  Tomita \etal (1996) failed to find the enhanced
fraction of blue galaxies in the region between these two X-ray peaks.

This paper aims at not only selecting faint member galaxies of \a168
by photometric redshift technique, but also investigating the
substructures in the spatial distribution and the line-of-sight
velocity distributions. We wish to provide the direct evidence of the
ongoing merge event of \a168 from our analysis of multicolor
photometries. This paper is structured as follows: In \S\ref{data}, we
describe the BATC observations, the data reduction. The technique of
photometric redshift is used to estimate the redshift for all galaxies,  
then the membership selection of \a168 is
performed.  In \S\ref{secsx}, the substructures of \a168 are explored
on the basis of the member galaxy distribution and
the X-ray emission of hot gas. The dynamics of substructures
is also studied. In \S\ref{seclfs} the luminosity
functions (LFs) of \a168 are analyzed.  The summary is given in
\S\ref{secsum}.

\section{DATA \label{data}}
\subsection{BATC Observation and Data Reduction \label{obs}}

The BATC program is a large-field ($58^{\prime} \times 58^{\prime}$)
multicolor sky survey based on the  60/90 cm f/3 Schmidt Telescope of
National Astronomical Observatories, Chinese Academy of Sciences
(NAOC). A Ford Aerospace 2048$\times$2048 CCD camera with the pixel
size of 15$\,\mu$m ( $\sim 1^{\prime\prime}.7/$pixel) is mounted on
the telescope. The BATC filter system contains 15 intermediate-band
filters (Fig.~\ref{figfilter})  covering a wavelength range from 3000
to 10\,000\,{\AA}, which are specifically designed for avoiding night
sky emission lines. A detailed description of the BATC photometric
system can be found in Fan \etal (1996) and Zhou \etal (2003).

\smallskip
\centerline{\framebox[14cm][c]{Fig. 1: Transmission of filters.}}
\smallskip

From August 1995 to March 2002, we totally accumulated about 37
hours exposure for \a168 with 13 filters. More information about our
observations is given in Table~\ref{obsparam}.  The raw data are
processed with an automatic data reduction software named PIPELINE I
(Fan \etal 1996) which includes the bias subtraction and the dome
flat-field correction.  The technique of integral pixel shifting is
used in the image combination during which the cosmic rays and bad
pixels are corrected by comparing multiple images.  The mean error of
position calibration is $\sim 0^{\prime\prime}.5$ by taking the Guide
Star Catalog (GSC) as standard.  By using the 
PIPELINE I$\!$I software
(Zhou \etal 2003), developed on the basis of the DAOPHOT kernel,
photometries are performed for all the objects detected.  As a result,
the photometries with the Point Spread Function (PSF) model and the
different apertures are obtained for 4798 objects detected in at least
3 bands. The flux calibration of the SEDs is performed by using the
Oke-Gunn standard stars which were observed during photometric nights.
The detailed information about calibration can be found in Zhou \etal
(2001).  Because we have no calibration image for $g$ filter, we
instead perform a model calibration that is developed specially for
the large-field photometric system by Zhou \etal (1999).

\smallskip
\centerline{\framebox[14cm][c]{Table. 1: The statistics of observations.}}
\smallskip

\subsection{Combining the BATC and the SDSS SEDs of galaxies}
\label{seccat}

\a168 has been observed by SDSS with five broad bands, namely
$u',g',r',i',z'$ (Fig.~\ref{figfilter}) , and the photometric data are distributed in the SDSS
Early Data Release (EDR, see Stoughton \etal 2002). The method of combining
the SDSS photometric data and the BATC SEDs is explored recently by
Yuan \etal (2003), and  this work shows that the combined SEDs 
could lead to a more accurate estimate of
photometric redshifts. We cross identify the BATC source list with the SDSS
photometric catalog in the same region, and a catalog containing 2580
galaxies and 2218 stars is achieved. The object classifications are
taken from the result of SDSS photometries. For the photometries of BATC, 
we adopt the PSF
magnitudes for stars, and take a fixed aperture with a radius of 4
pixels (\ie $r_{\rm ap} \sim 6^{\prime\prime}.8 $) for galaxies. In
order to combine the photometric results of the BATC and the SDSS, an
aperture-correction should be applied to the SDSS model magnitudes
($m_{\rm model}$) by \beq \Delta m = m_{\rm ap} - m_{\rm model} =
-2.5\,\log \frac{\int^{r_{\rm ap}}_{0}2\pi r I(r){\rm d}r}
{\int^{\infty}_{0}2\pi r I(r){\rm d}r}, \eeq where $m_{\rm ap}$ is the
aperture magnitude, $I(r)$ is the preferred profile function of
surface intensity (\eg the power law or de Vaucouleurs $r^{1/4}$
model). The model magnitude in each band and its corresponding
parameters that are used to quantify the preferred brightness profile
can be found in the SDSS data.

To observe the possible systematic offset between the BATC and the
SDSS photometric system, we derive the zero point difference of 
SED (ZPDS) for each
object by calculating the difference between the SDSS $r'$ magnitude
and the interpolated BATC magnitude at 6180\AA. Fig.~\ref{figdbs}
shows the distribution of ZPDSs for all the objects
detected. For stars, the magnitude offset between these two
photometric systems approximates to zero, indicating  that the BATC
and SDSS photometric systems are in good agreement.  For galaxies, a
more widely spread distribution of ZPDSs is found. 
This might be the result from the errors in photometries and the
uncertainties in model fittings. For adapting to BATC multicolor
photometric system, the SDSS SEDs are corrected by
adding their ZPDS. Finally, we obtain the combined SEDs,
including 13 BATC bands plus 5 SDSS bands, for 1553 galaxies brighter
than $r'=20$ mag.

\smallskip
\centerline{\framebox[14cm][c]{Fig. 2:  Zero point distribution.}}
\smallskip

\subsection{Photometric Redshift and Membership Determination}
\label{secmem}

The photometric redshift technique was originally developed for detecting
the high-$z$ objects based on the broad-band photometries
(Pell{\' o} \etal 1999; Bolzonella, Miralles, \& Pell{\' o} 2000). 
The 15-intermediate-band SEDs obtained by the BATC photometric system
can be regarded as the rough
spectrum. The advantages of the BATC data in measuring the photometric
redshift of some faint galaxies can be expected.  With the BATC SEDs
(Yuan \etal 2001, Xia \etal 2002) and the combined SEDs of the BATC
and the SDSS (Yuan \etal 2003), the photometric redshifts of
faint galaxies can be estimated with a certain accuracy ($\Delta
z_{\rm phot}\approx 0.02$), which allows us to isolate the member
galaxies for some nearby galaxy clusters.

The technique of photometric redshift has been applied to the 18-band
SEDs with the help of the {\sl Hyperz} code (Bolzonella, Miralles, \& Pell{\' o}
2000). The red-shifted spectral templates of normal galaxies are used
to fit the observed SEDs, and the reddening law is set to be a free
parameter during the  fitting.  The photometric redshift for each
galaxy is searched in a redshift range from 0.0 to 1.0, with a
searching step of 0.005.  The histogram of $z_{\rm phot}$ for all
galaxies is plotted in Fig.~\ref{figstg}, and most of them are located
at $z<0.5$. 195 galaxies with known spectroscopic redshifts $z_{\rm
spec} < 0.25$ are selected to derive the uncertainty of photometric
redshift which is found to be $\sim$ 0.024 (see Fig.~\ref{figsig}).

\smallskip
\centerline{\framebox[14cm][c]{Fig. 3:  The histogram of photometric redshift. }}
\smallskip

\smallskip
\centerline{\framebox[14cm][c]{Fig. 4: A comparison of photometric redshifts and spectroscopic redshifts.}}
\smallskip

According to the known spectroscopic redshifts,  there are 121
galaxies distributed in a range from 0.04 to 0.05. We take them as the
member galaxies of \a168. About 90\% (109/121) galaxies have
photometric redshifts within the range $[0.02,0.07]$, thus we take
this range as the criterion for the membership determination based on the
photometric redshift. As a result,
255 galaxies are selected as the member candidates. Together with the
spectroscopically confirmed members, we totally obtain an enlarged
sample of 376 member galaxies. It should be noted that the
uncertainty of photometric redshift is rather large compared with the
intrinsic dispersion of \a168, but it still can be used to select the
member galaxy candidates with high efficiency, especially for the
faint galaxies.

\section{SUBSTRUCTURES \label{secsx}}
\subsection{Spacial Distribution of Members and X-ray Map \label{secsp}}
Fig.~\ref{figsp} gives the spatial distribution of 376 member galaxies
(small circles), with its corresponding contour map (thin lines) of
the number density superposed. 
In the figure, D indicates the density peak located at
${01^{\rm h}15^{\rm m}06^{\rm
s}}$~+$00^\circ19^{\prime}12^{\prime\prime}$ (J2000.0).  The spatial
distribution is elongated along the NW-SE direction with a position angle
of $\sim 30^{\circ}$.  A possible group (A168WG), indicated by E 
in the figure, can be recognized to the west of the
cluster core.  The group is close to the high density region and seems
to be falling down to the center, increasing the galaxy density
there. A similar spatial distribution was earlier derived by
Kriessler \& Beers (1997), but their sample of 106 galaxies did
not show the core structures and the group clearly.
We take the density peak as the center, and then derive the radial
number density profile in Fig.~\ref{figpro}. It is clear that most of the 
galaxies concentrate within a range of about 12\arcmin  
($\sim$~0.7~\mpc).

\smallskip
\centerline{\framebox[14cm][c]{Fig. 5: The member distribution and the X-ray map.}}
\smallskip

\smallskip
\centerline{\framebox[14cm][c]{Fig. 6: The number density profile.}}
\smallskip

The X-ray image in $0.2-3.5$~keV band, 
observed by IPC of Einstein Observatory, is overlayed on the
spatial distribution in Fig.~\ref{figsp} (thick lines). It is
interesting that the X-ray map and the member distribution coincide
well with each other both in their spatial locations and their shapes.
As mentioned above (\S\ref{secintr}), the X-ray emission of Abell~168
has two peaks, marked by A and B in Fig.~\ref{figsp}.  The higher
X-ray peak, point A, is located at ${01^{\rm h}14^{\rm m}57^{\rm
s}}$~+$00^\circ25^{\prime}04^{\prime\prime}$ (J2000.0) with a distance of
$\sim 6^{\prime}$ (\ie $\sim$ 320 kpc) to the density peak.
In the figure, C points to the cD galaxy UGC~00797.  Point A is very
close to the cD, which agrees with the traditional experience that D/cD
galaxies are located around the peak of the X-ray emission 
(Jones \etal 1979; Forman \& Jones 1982).  
However, this cD galaxy  significantly deviates from
the local density peak of Abell~168, which seems inconsistent with the
idea that a D/cD appears to be associated with the center of 
its host cluster 
(Beers \& Geller 1983; Oegerle \&
Hill 2001).  On the other hand, the lower X-ray peak, point B, is
surrounded by the clustering bright galaxies.  From the shape of the
X-ray emission, the hot gas between the two peaks seems to be pushed
away from the center by the head-on collision of the two clumps.
Combining all above characteristics, it seems that there are some
substructures in \a168.  Ulmer, Wirth \& Kowalski (1992) pointed out
that Abell 168 was possibly formed  by the collision of two
subclusters. If this is true, it is reasonable to assume 
the collision occurs along the elongated direction.  To explore the
substructures in detail,  we artificially divide the main region of
Abell~168 (indicated by a $0.4^{\circ}\times0.6^{\circ}$ box in
Fig.~\ref{figsp}) into two equal parts. One is centered at the higher
X-ray peak with 79 galaxies. We call it north subcluster (A168N). Another
is located at the lower X-ray peak with 97 galaxies, named south
subcluster (A168S).  In the following sections we try to find other
evidences to support that they are indeed physically different parts 
in Abell 168, in other words,
they are really two subclusters, and we will show that they seem to be
under a merging process.

\subsection{Dynamics of Substructures}
Provided Abell 168 is formed by merging of two subclusters,  the
dynamics  remaining after the collision should be presented by testing
the velocity distribution. On the basis of the spectroscopically 
confirmed member galaxies, we are allowed to study the dynamics of \a168 in detail. 
The statistical parameters of A168N, A168S, A168WG and A168 are listed in Table~\ref{tbvst}. The mean redshift ($z$), the mean velocity 
($v_{\rm r}$) and the velocity dispersion ($\sigma_{\rm r}$) are all calculated by the biweight statistic in ROSTAT software 
(Beers, Flynn, \& Gebhardt 1990), except the velocity dispersion of A168WG is taken 
from the gapping statistic.
The asymmetry index (AI) and the tail index (TI) of the velocity distributions
are also provided by ROSTAT.
Fig.~\ref{figvs} 
shows the velocity distribution of these two subclusters, respectively. The distribution
of A168N is more
concentrated than that of A168S. It seems that, for A168S, there is a gap 
at $\sim 13\,000$ \kms, which can be clearly seen from a stripe
density plot in Fig.~\ref{figstripe}. 
According to the value of TI (see Bird \& Beers 1993 for detail), A168N possesses a light-tailed 
distribution, while the distribution of A168S is near Gaussian.
This suggests that these two distributions are intrinsically different.
Moreover, the classical K-S test also shows the velocity distributions of these
two subclusters are from the different parent distributions with 94\% 
probability. 

\smallskip
\centerline{\framebox[14cm][c]{Fig. 7: $V_{\rm r}$ distribution of
A168N and A168S.}}
\smallskip

\smallskip
\centerline{\framebox[14cm][c]{Fig. 8: Stripe density plot.}}
\smallskip

Although it appears the obvious interaction between these two subcluster 
according
to the spatial distribution, one should keep in mind 
this is a projection effect. So 
it is necessary 
to estimate the probability that they are gravitationally bound, in order to 
confirm they are really two interacting subclusters. We take a
linear two-body model for the system as Beers, Geller, \& Huchra (1982) 
did in analyzing the cluster 
Abell 98. Under this model, the relative velocity $V$ and the separation 
$R$ between two subclusters can be converted to the observational quantities: 
the relative radial velocity $V_{\rm r}=V \sin \alpha$ and the projection separation
$R_{\rm p}=R\cos\alpha$. The projection angle 
$\alpha$ ($\in [0,\pi/2]$)
is defined as the the angle between the line connecting the two subclusters and
the plane of sky. The energy relation for gravitational binding is 
taken from the Newtonian criterion
\beq
\label{eqb}
\frac{V^2_{\rm r}R_{\rm p}}{2GM}\leq \sin^2\alpha \cos \alpha,  
\eeq
where $G$ is the gravitational constant. $M$ is the system mass, equal to the total mass of the two subclusters.
The difference of mean velocity between A168N and A168S can be easily 
figured out $V_{\rm r}=264\pm 142$ \kms. We take $R_{\rm p}=0.584$ \mpc, 
the distance between the two X-ray peaks which roughly represent the centers
of these two subclusters. 
Following the method described in Beers, Geller, \& Huchra (1982), the mass
and the gravitational scale lengths of A168N, A168S and A168 are estimated and listed in Table~\ref{tbvst}. 
The probability of gravitational binding can be  
obtained by the formula 
\beq
P=\frac{1}{A}\int_{0}^{+\infty}p(V_{\rm r})\,p(\alpha|V_{\rm r})\,{\rm d}V_{\rm r},
\eeq
where $A$ is re-normalization factor. 
$p(V_{\rm r})$ is the probability distribution of the velocity which is 
assumed to be a Gaussian. $p(\alpha|V_{\rm r})$, which can be obtained 
from Eq.~\ref{eqb}, is the probability of 
the valid $\alpha$ for gravitational binding at a given $V_{\rm r}$. Integration is over all the 
appropriate values of velocity.
As the results, we find these two 
subclusters are gravitationally bound with a probability of 92\%.

Another efficient statistic method to test the substructures of the radial
velocity distribution was recently introduced by Colless \& Dunn
(1996), who keep the spirit of Dressler-Shectman test (Dressler \&
Shectman 1988) and employ the standard K-S test to compare the
velocity distribution of the local group, $n$ nearest neighbors, and
the whole velocity distribution. The statistic is  defined as \beq
\kappa_n=\sum^{N}_{i=1}-{\rm log}~[P_{\rm KS}(D>D_{\rm obs})], \eeq
where $N$ is the number of all  member galaxies of a cluster,  and $D$
is the statistic in standard K-S test.  For each galaxy, we compute the
probability,  $P_{\rm KS}(D>D_{\rm obs})$, of its local group via
Monte Carlo simulation by randomly shuffling velocities every time.
This probability gives the an  indication of how significant the
observed $D_{\rm obs}$ is by comparing with the simulations.
$\kappa_n$ is an indication that the local velocity is different from
the overall distribution.

We take the group size of $n=10$. By involving  $10^5$ simulations, we
find that  $\kappa_{10}=0.000\%$,  almost no simulated $D$ greater
than the observed $D_{\rm obs}$.   A significant deviation of local
velocity distribution from the overall distribution can be found in
Fig.~\ref{bub}. The bubble radius for each galaxy is proportional to
$-{\rm log}~[P_{\rm KS}(D>D_{\rm obs})]$.  
Consistent with the result of spectroscopic redshifts, no
further substructure was found in the central region,
because the photometric redshift is
not precise enough to distinguish the small difference, 
such as $\sim$~260~\kms, in velocity.  
The result possibly suggests
that the two subclusters are approaching each other along the plane
perpendicular to our line of sight.  The obvious bubble concentration
at the west possibly supports the existence of the west group
discussed in the section \S\ref{secsp}.

\smallskip
\centerline{\framebox[14cm][c]{Fig. 9: Bubble plot.}}
\smallskip

\section{LUMINOSITY FUNCTIONS}\label{seclfs}
\subsection{{\sl Double} Schechter Function }
\label{secdsf}
The luminosity function is a key diagnostics for clusters of galaxies
because it is tightly related to the dynamical evolution and the
merging history  of galaxy clusters.  In the past decades, the LFs of
GCs are well described by the famous Schechter function, which was
originally proposed by Schechter (1976):
\begin{equation}
\label{eqSF}
\phi(L)dL=\phi^*\,(L/L^*)^\alpha\,\exp(-L/L^*)\,{\rm d}(L/L^*),
\end{equation}
where $\phi^*, L^*, \alpha$ are the normalization parameter, the
characteristic luminosity and the faint slope parameter, respectively.

Recently, the deep observation of the Coma cluster
reveals their LFs possess an enhanced faint tail (\eg Trentham \&
Tully 2002). It is widely perceived that a single Schechter function
gives a poor representation of the data when the LFs extend to the
fainter end (Biviano \etal 1995; Durret, Gerbal, Lobo \& Pichon 1999;
Trentham \& Tully 2002).  The segregation in LFs of GCs suggests that
there are at least two populations of the galaxies in clusters (Durret,
Adami \& Lobo 2002).  For example, Ferguson \& Sandage (1991) take a
form of Gaussian part plus Schechter function in order to probe the
two populations, the giants and the dwarfs.

The Schechter function works well for the bright end as shown in the
previous literatures, such as Dressler (1978), Paolillo \etal (2001). 
However, the present extending LFs require a variation to
describe the enhanced tail rather than a single Schechter function.
Hence we propose an acceptable {\sl double} Schechter function (DSF
hereafter).
\begin{eqnarray}
\label{eqdsf}
N(M) & = \nonumber& \phi^*_0\,[10^{-0.4(M-M^*_0)}]^{\alpha_0}\,{\rm
     e}^{10^{-0.4(M-M^*_0)}} \\ & + &
     \phi^*_1\,[10^{-0.4(M-M^*_1)}]^{\alpha_1}\,{\rm
     e}^{10^{-0.4(M-M^*_1)}}.
\end{eqnarray}
Each term has the same form as the Schechter function in
Eq.~\ref{eqSF}.  $M^*_i\,(i=0,1)$ is the characteristic absolute
magnitude with an equivalent characteristic luminosity;
$\phi^*_i,\alpha_i,(i=0,1)$ have the same meanings as in Eq.~\ref{eqSF}.

\subsection{Luminosity Functions}
\label{secLF}
We use DSF to fit the observed LFs by means of the $\chi^2$
minimization. The results are listed in Table~\ref{tbdsffit} and
plotted in Fig.~\ref{figlum}. A168S shows a decay tail while A168N shows an
increasing one, suggesting that the galaxy contents of these two
regions are quite different.  We also compute the ratio of bright
galaxies ($M_{r^{\prime}}\leq -18.5$)  to the faint
($M_{r^{\prime}}>-18.5$) for these two subclusters.  This ratio for
A168N is found to be 32/47=0.68.  However, for A168S, the
ratio is 56/41=1.37.  It is obvious that the bright galaxies are more
abundant in A168S while the faint galaxies prefer to occur in A168N.
This can be interpreted by the cannibalism model (Hausman \& Ostriker
1978). That is to say: the cDs are likely to be formed by accreting the
surrounding massive galaxies, which naturally results  in the deficit of
bright galaxies in the most dynamically evolved clusters (Lugger
1986).  Hence, the formation and evolution of these two subclusters might be
different. A168N, associated with the cD galaxy UGC~00797,  is
likely to be more evolved than A168S.  To confirm our dividing of 
A168N and A168S is not accidental. We try other methods of 
dividing the core region. For example, we divide the core region into 
an east  and a west part equally along the elongated direction of 
the spatial distribution. In this case, the clear contrast in the LFs 
disappears,
which supports the dividing of A168N and A168S is a special choice, 
and they are the two subclusters under the
merging event of \a168.

\smallskip
\centerline{\framebox[14cm][c]{Table. 2: The fitting results of LFs of
\a168.}}
\smallskip

\smallskip
\centerline{\framebox[14cm][c]{Fig. 10:The observed luminosity
functions are fitted by DSF. }}
\smallskip

Panel (c) in Fig.~\ref{figlum} shows the LF of the  central region
(A168N+S) of the cluster. As shown in previous works of Dressler (1978) and
Oegerle, Ernst, \& Hoessel (1986), the LF has a quite flat tail which
is, however, an average of  LFs of A168N and  A168S according to the present
work. As for the whole field, the observed LF and its fitting model
are shown  in panel (d) of Fig.~\ref{figlum}.  For the overall LF,
DSF gives a good description to the obvious enhanced faint tail at
$M_{r^{\prime}}>-18.5$ mag.  Compared with the LF of the central region,
the overall LF suggests that the faint galaxies (possible dwarfs) tend
to be located in the outer region of the cluster.  
This perhaps support the
population-density relation: dwarfs are more common in lower density
environments (Phillipps, Driver, Couch, \& Smith 1998).

\section{SUMMARY}
\label{secsum}

We accumulated about 37 hours of high quality observations of Abell
168 with 13 intermediate filters of BATC photometric system.  In the
EDR of SDSS, we find that the photometry and the
spectroscopic redshift measurement are completed in the same field. After a
cross-identification with SDSS, we  combine the photometric data of the objects
detected in two systems getting the 18-band SEDs, which include 13
BATC bands and 5 SDSS bands.  A complete sample of 1553 galaxies brighter
than $r^{\prime}=20$ mag  is achieved for the subsequent study.

By using the technique of photometric redshift, we get an enlarged
sample of 376 galaxies, including 121 spectroscopically
identified member galaxies and 255 newly selected member candidates.
Spatially, the member galaxies show an elongated distribution along
the NW-SE direction with a position angle of $\sim 30^{\circ}$, which
is consistent with the spatial characteristic of the X-ray image  obtained
by Einstein Observatory.  Moreover, the two X-ray peaks and the
deviation of the cD galaxy UGC~00797 from the local density peak support the
existence of substructures in \a168. According to the spatial
characteristics, we artificially divide the core region into two
parts, A168N and A168S.  
By employing ROSTAT software, a detailed investigation of the dynamics
of these two subclusters is performed. A relative radial velocity of 
$264\pm142$ \kms was found between them. 
Their velocity distributions are intrinsically
different and should be from the different parent distributions. 
Via building a linear two-body model, 
they are found to be gravitationally bound with a probability of 92\%. 
Investigation of LFs by
the DSF shows an obvious difference in the galaxy content of 
these two subclusters.  
The A168S is dominated by bright galaxies while A168N has
larger number of faint galaxies.  A168N is a sub-system  associated
with the cD galaxy and likely to  be a more evolved
system than A168S.  All these indications are consistent and support that
\a168 is likely to be a head-on merging system at its early stage of
the process,  and that the collision plausibly occurs along the direction
perpendicular to our line of sight.

\acknowledgments  The authors would like to thank the referee Timothy
Beers who gives the good suggestions to improve the paper.  We are
grateful to the following persons for their valuable suggestions and
discussion: Dr. HaiGuang Xu, SuiJian Xue,  XiangPing Wu,
XiaoFeng Wang,  Yu Liu, ZhengYu Wu, Mr.\ Yu Lu, Ms.\ Bin Yang,  LiFang
Xia,   and especially Mr.\ Albrecht R{\"u}diger.  We also appreciate
the assistants who contributed their hard work to the observations.
This research has made use of the NASA/IPAC Extragalactic Database
(NED), the EDR of SDSS, High Energy Astrophysics Science Archive
Research Center (HEASARC).  This work is supported by the National Key
Base Sciences Research Foundation (NKBRSF, TG199075402) 
and is also supported by the
Chinese National Science Foundation (NSFC).

\clearpage

\begin{table}[ht]
{\footnotesize
\caption{\label{obsparam}Parameters of the BATC filters and the
statistics of observations of \a168.}
\vspace {0.5cm}
\begin{tabular}{cccccccccc}
\tableline \tableline No.  & Filter& $\lambda_{\mbox{eff}}$ &FWHM & Exposure&
N$_1$\tablenotemark{a} & Seeing\tablenotemark{b} &
N$_2$\tablenotemark{c} \\ 
&   & (\AA) &(\AA)&(hours) & &(arcsec)\\ 
(1) & (2)& (3)& (4) & (5)& (6) & (7)& (8)\\ \tableline 
 1 & $c$& 4210& 309 & 3.67& 11 &5.08& 1\\ 
 2 & $d$& 4546& 332 & 5.33& 16 &4.09& 1\\
 3 & $e$& 4872& 374 & 4.00& 12 &3.94& 2\\ 
 4 & $f$& 5250& 344 & 3.47& 11 &4.18& 4\\ 
 5 & $g$& 5785& 289 & 1.33&\ 4 &3.81& 0\\ 
 6 & $h$& 6075& 308 & 1.00&\ 3 &4.80& 3\\ 
 7 & $i$& 6710& 491 & 1.08&\ 4 &3.80& 3\\ 
 8 & $j$& 7010& 238 & 2.00&\ 6 &3.94& 1\\ 
 9 & $k$& 7530& 192 & 2.00&\ 6 &4.49& 3\\
10 & $m$& 8000& 255 & 5.67& 17 &3.94& 1\\ 
11 & $n$& 8510& 167 & 2.67&\ 8 &4.81& 2\\ 
12 & $o$& 9170& 247 & 1.33&\ 4 &6.29& 1\\ 
13 & $p$& 9720& 275 & 3.40& 11 &5.03& 2\\ \tableline
\end{tabular}\\
\tablenotetext{\rm a}{Number of images of each color.}
\tablenotetext{\rm b}{The seeing is of the combined images.}
\tablenotetext{\rm c}{Number of the calibration images.}
}
\end{table}

\clearpage
\begin{table}[ht]
{\footnotesize
\caption{\label{tbvst}Parameters of \a168.\tablenotemark{\dag}}
\vspace {0.5cm}
\begin{tabular}{l|r@{}lr@{}lr@{}lr@{}l}
\tableline \tableline 
Parameter 
& \multicolumn{2}{c}{A168N} 
& \multicolumn{2}{c}{A168S}  
& \multicolumn{2}{c}{A168WG} 
& \multicolumn{2}{c}{A168} \\ \tableline
$N_{\rm gal}$\tablenotemark{a}
& \multicolumn{2}{c}{79} 
& \multicolumn{2}{c}{97}  
& \multicolumn{2}{c}{19} 
& \multicolumn{2}{c}{376} \\
$N_{\rm gal,\ spec}$\tablenotemark{b}
& \multicolumn{2}{c}{26} 
& \multicolumn{2}{c}{52}  
& \multicolumn{2}{c}{6} 
& \multicolumn{2}{c}{121} \\
$z$ 
&$0.0446$&$\ \pm\ 0.0004$
&$0.0455$&$\ \pm\ 0.0003$
&$0.0466$&$\ \pm\ 0.0001$  
&$0.0451$&$\ \pm\ 0.0002$ \\
$v_{\rm r}$ (\kms)
&$ 13065 $&$\ \pm\   113$ 
&$ 13329 $&$\ \pm\   86 $
&$ 13633 $&$\ \pm\   35 $
&$ 13206 $&$\ \pm\   50 $ \\
$\sigma_{\rm r}$ (\kms) 
&$ 564 $&$\ \pm\     90 $
&$ 613 $&$\ \pm\     56 $
&$ 491 $&$\ \pm\     189$
&$ 554 $&$\ \pm\     34 $ \\     
$\langle 1/r_{\rm p}\rangle^{-1}$ (\mpc)\tablenotemark{c}
& \multicolumn{2}{c}{0.30} 
& \multicolumn{2}{c}{0.30}  
& \multicolumn{2}{c}{---} 
& \multicolumn{2}{c}{0.77} \\
$M$ ($10^{14}$\,\Msun)
& $2.1  $&$\ \pm\    0.7$ 
& $2.5  $&$\ \pm\    0.5$
& \multicolumn{2}{c}{---} 
& $5.2  $&$\ \pm\    0.6$ \\      
AI\tablenotemark{d}
& \multicolumn{2}{c}{$-0.263$} 
& \multicolumn{2}{c}{$-0.392$}  
& \multicolumn{2}{c}{$-0.327$} 
& \multicolumn{2}{c}{$-0.093$} \\
TI\tablenotemark{e}
& \multicolumn{2}{c}{$\ \ 1.458$} 
& \multicolumn{2}{c}{$\ \ 0.958$}  
& \multicolumn{2}{c}{$\ \ 0.000$} 
& \multicolumn{2}{c}{$\ \ 0.901$} \\
\tableline
\end{tabular}\\
\tablenotetext{\dag}{The errors are computed at 68\% confidence level.}
\tablenotetext{\rm a}{The number of galaxies including the
spectroscopically confirmed members and the photometric redshift
selected candidates.}  \tablenotetext{\rm b}{The number of galaxies only
including the spectroscopically confirmed members.}
\tablenotetext{\rm c}{The gravitational scale length.} 
\tablenotetext{\rm d}{Asymmetry index.}  \tablenotetext{\rm e}{Tail index.}  }
\end{table}

\clearpage
\begin{table}[ht]
{\footnotesize
\caption{\label{tbdsffit}The fitting results of LFs of \a168.
\tablenotemark{\dag}}
\vspace {0.5cm}
\begin{tabular}{lrrrrrrr}
\tableline \tableline &$\phi^*_{0}$ &$M^{*}_{0}$ & $\alpha_{0}$ &$\phi^*_{1}$ &
 $M^{*}_{1}$ & $\alpha_{1}$ & $\chi^2_{\rm min}$ \\ \tableline 
A168N 
&   14.9$^{+    3.4}_{   -2.9}$
& $-$21.75$^{+   0.67}_{  -2.50}$
&   0.04$^{+   0.08}_{  -0.07}$
&   37.5$^{+   20.2}_{  -15.9}$
& $-$16.91$^{+   0.56}_{  -0.63}$
&   0.40$^{+   0.98}_{  -1.00}$
&    1.6\\
A168S 
&   54.4$^{+    8.5}_{   -7.4}$
& $-$20.47$^{+   0.31}_{  -0.40}$
&   0.36$^{+   0.08}_{  -0.08}$
&  0.00$^{}_{}$
&  ---  $^{}_{}$
&  ---  $^{}_{}$
&    3.4\\
A168N+S 
&   50.3$^{+    5.6}_{   -5.2}$
& $-$21.14$^{+   0.32}_{  -0.48}$
&   0.05$^{+   0.04}_{  -0.04}$
&  0.00$^{}_{}$
&  ---  $^{}_{}$
&  ---  $^{}_{}$
&    4.1\\
Total 
&   95.8$^{+   10.}_{   -9.4}$
& $-$20.77$^{+   0.23}_{  -0.30}$
&   0.17$^{+   0.05}_{  -0.05}$
&  138.9$^{+   24.5}_{  -22.6}$
& $-$17.48$^{+   0.19}_{  -0.20}$
&  $-$0.19$^{+   0.28}_{  -0.26}$
&    3.6\\
\tableline
\end{tabular}\\
\tablenotetext{\dag}{The errors are computed at 68\% confidence level.}
}
\end{table}

\clearpage

\begin{figure}
\epsscale{0.5}
\plotone{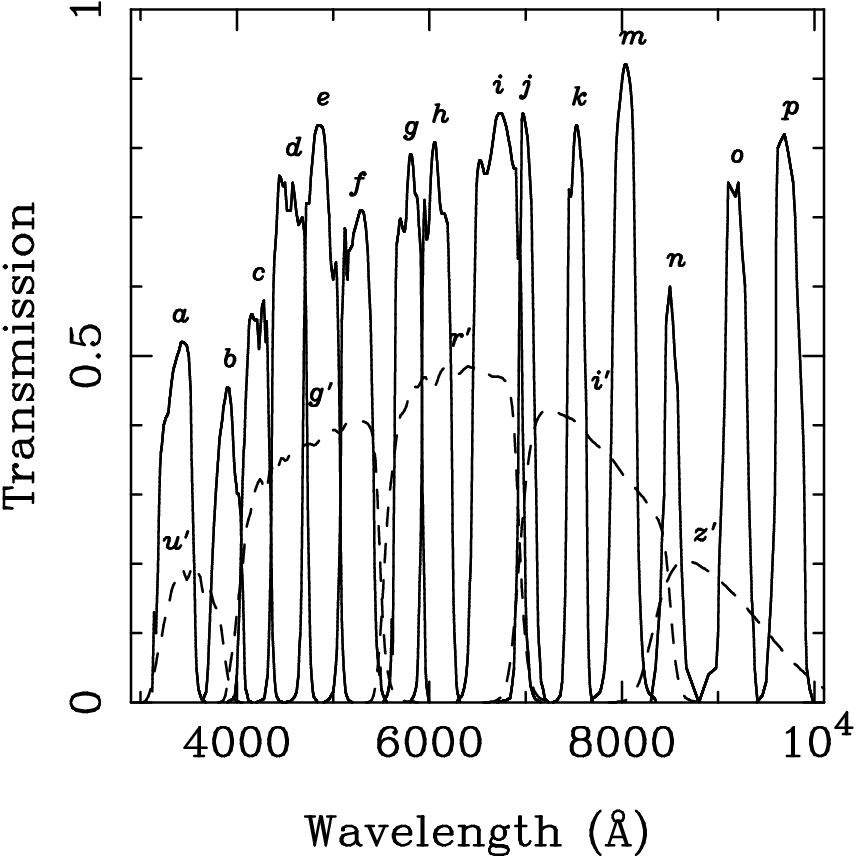}
\caption{
The transmission curves of filters in the BATC (solid lines) and 
the SDSS (dashed lines) photometric system. The names of the filters
are marked at the top of each curve. \label{figfilter} }
\end{figure}

\clearpage

\begin{figure}
\epsscale{0.5}
\plotone{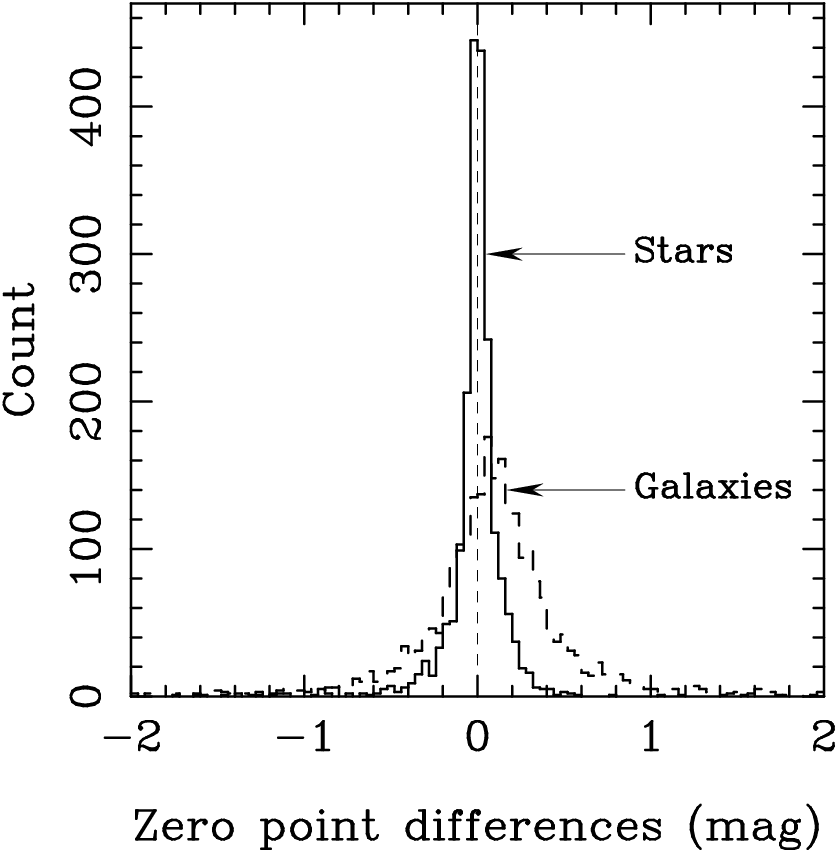}
\caption{The histogram of the zero point differences of 
individual SEDs in the BATC and the SDSS system 
for all galaxies and stars. \label{figdbs}}
\end{figure}

\clearpage
\begin{figure}
\epsscale{0.8}
\plotone{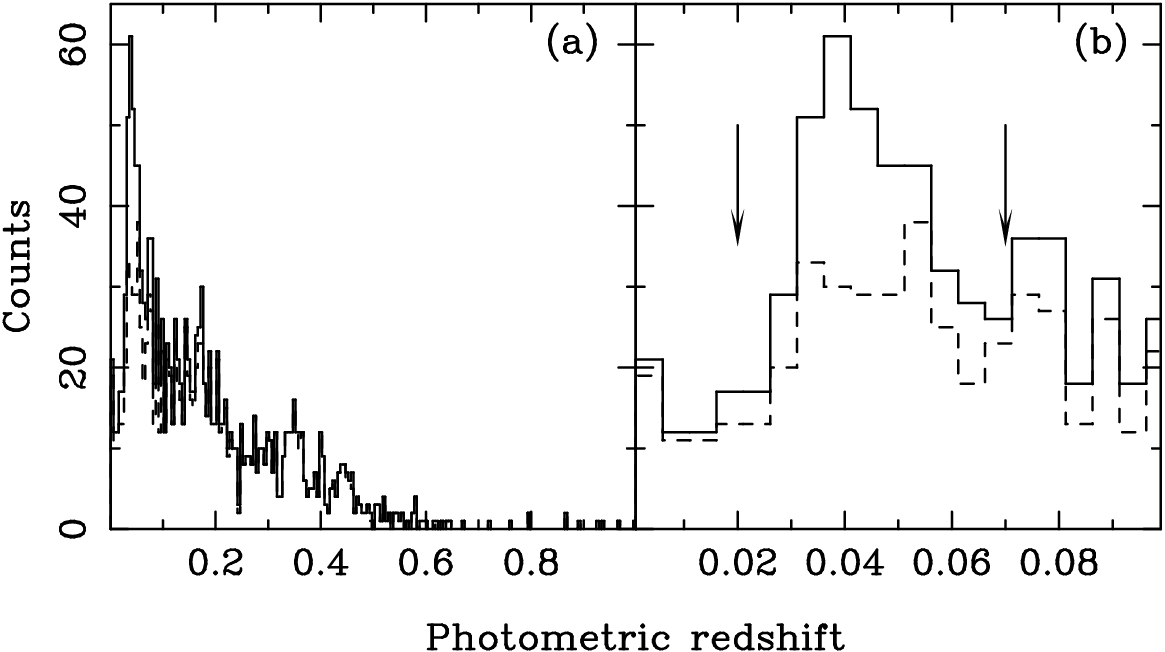}
\caption{Panel (a) shows the histogram of photometric redshift for
1553 galaxies.  The distribution from 0.0 to 0.1 is shown enlarged in panel
(b). In both figures, the solid lines represent photometric redshifts
of all galaxies, the dashed lines indicate photometric redshifts of
the galaxies without spectroscopic redshift. The range of [0.02,~0.07]
marked by two arrows is taken as the criterion for member candidates.
\label{figstg}}
\end{figure}

\clearpage
\begin{figure}
\epsscale{0.5}
\plotone{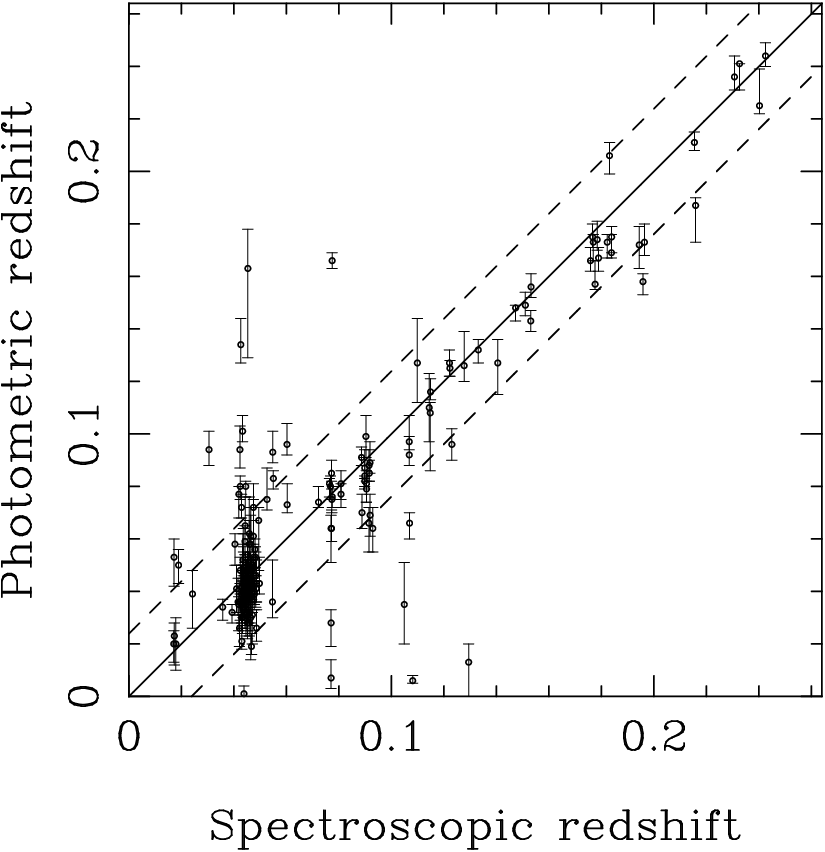}
\caption{A comparison of photometric redshifts and spectroscopic
redshifts for 195 known galaxies. The dashed lines mark out the
two-sigma region of the photometric redshifts.\label{figsig} }
\end{figure}

\clearpage
\begin{figure}
\epsscale{0.8}
\plotone{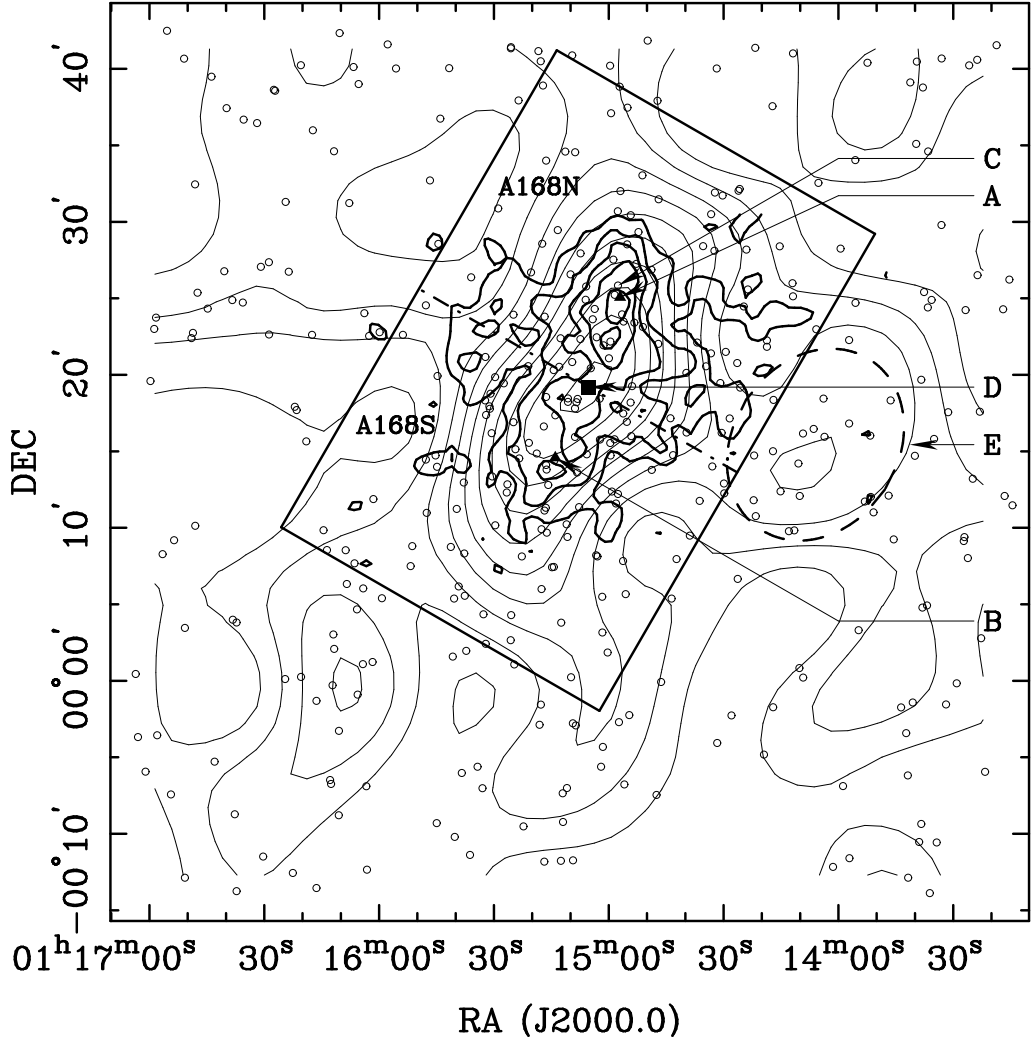}
\caption{Member distribution (small circles) of \a168, with 
corresponding contour map superimposed (thin lines).
Point A and point B are
the two X-ray peaks. C points to the cD galaxy UGC 00797. D points
to the number density peak. E points to the possible west group indicated by
a dashed line ellipse.  The box
($0.4^{\circ}\times0.6^{\circ}$) indicates the central region of the
cluster. The dotted-dashed line divides the north subcluster (A168N) and
the south subcluster (A168S) equally (see \S\ref{secsx} for detailed
analysis). \label{figsp} }
\end{figure}
\clearpage

\begin{figure}
\epsscale{0.6}
\plotone{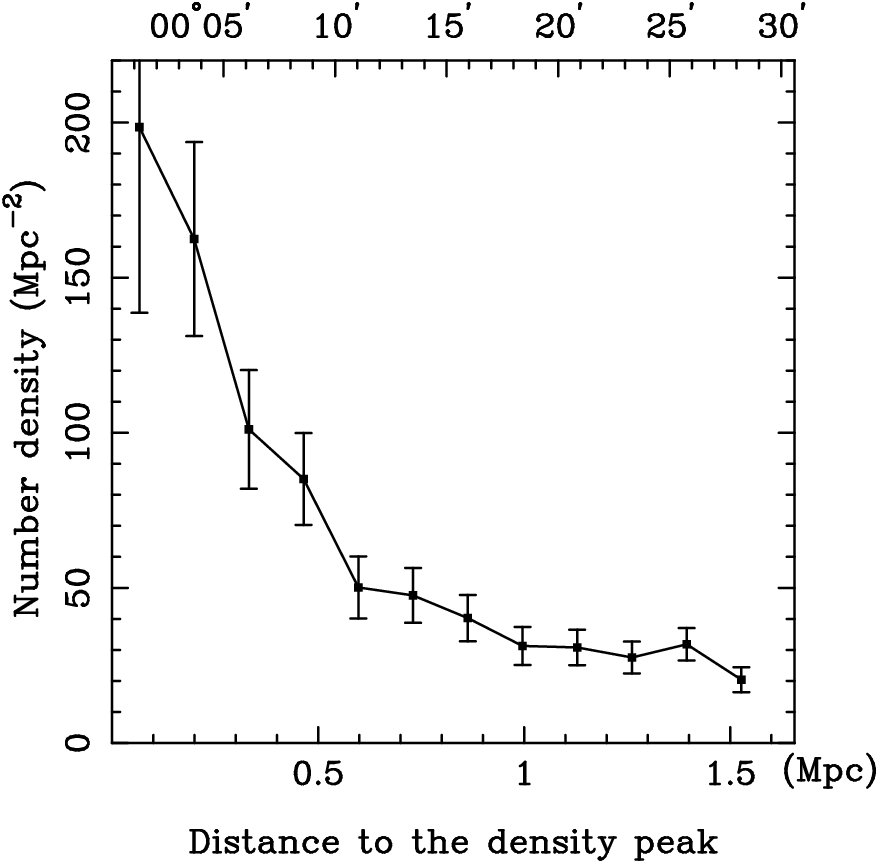}
\caption{The radial number density profile. The bottom abscissa shows the distance in unit of Mpc.  The top abscissa is in unit of arcmin.\label{figpro} }
\end{figure}
\clearpage

\begin{figure}
\epsscale{0.6}
\plotone{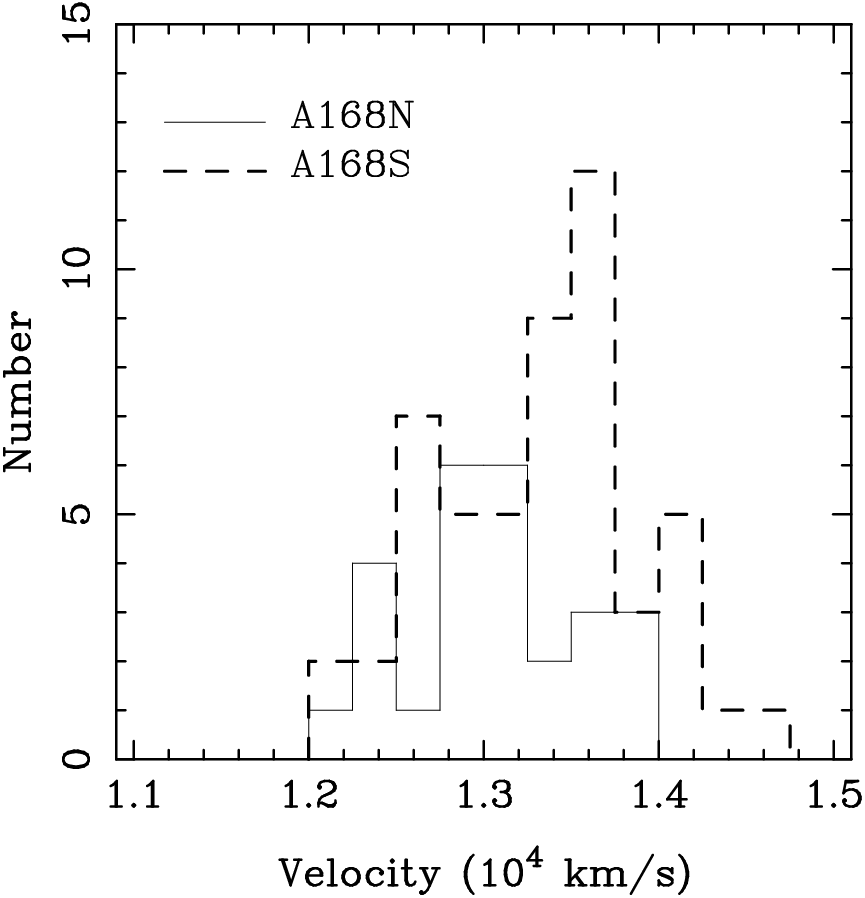}
\caption{The velocity distributions of the spectroscopically confirmed 
member galaxies for A168N and A168S are presented by the solid line and the dashed line, respectively.  \label{figvs} }
\end{figure}
\clearpage

\begin{figure}
\epsscale{0.6}
\plotone{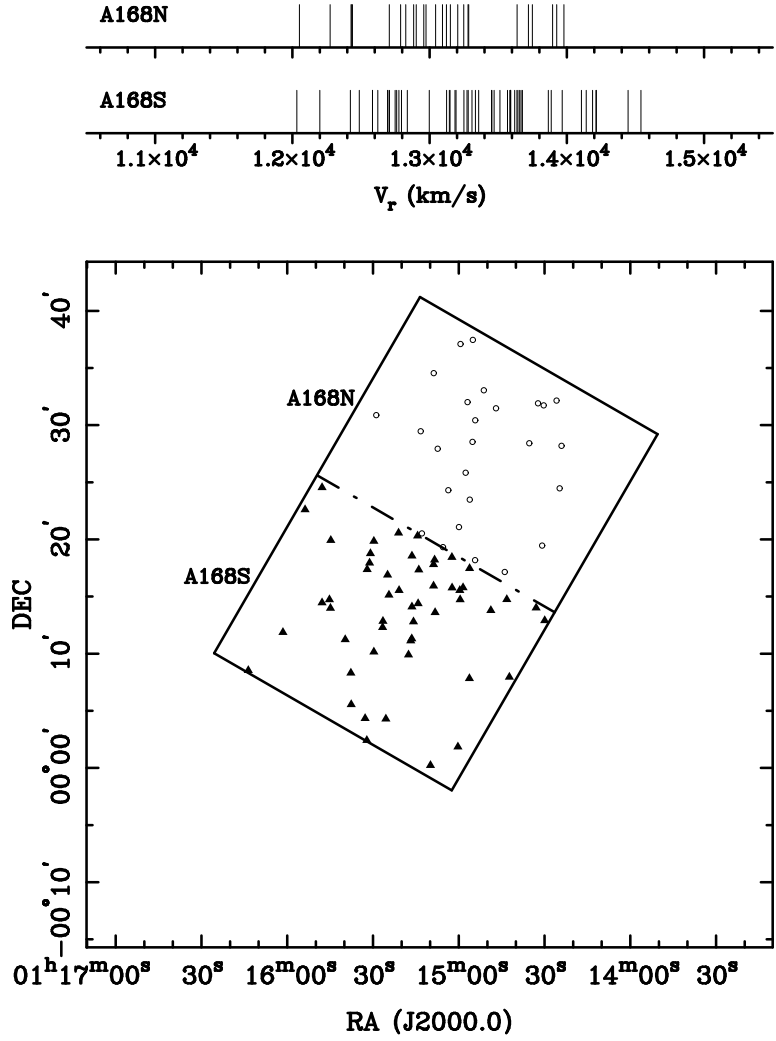}
\caption{For the spectroscopically confirmed members in these two subclusters, the stripe density plot of velocity is plotted above the spatial distribution.
\label{figstripe} }
\end{figure}

\clearpage

\begin{figure}
\epsscale{0.8}
\plotone{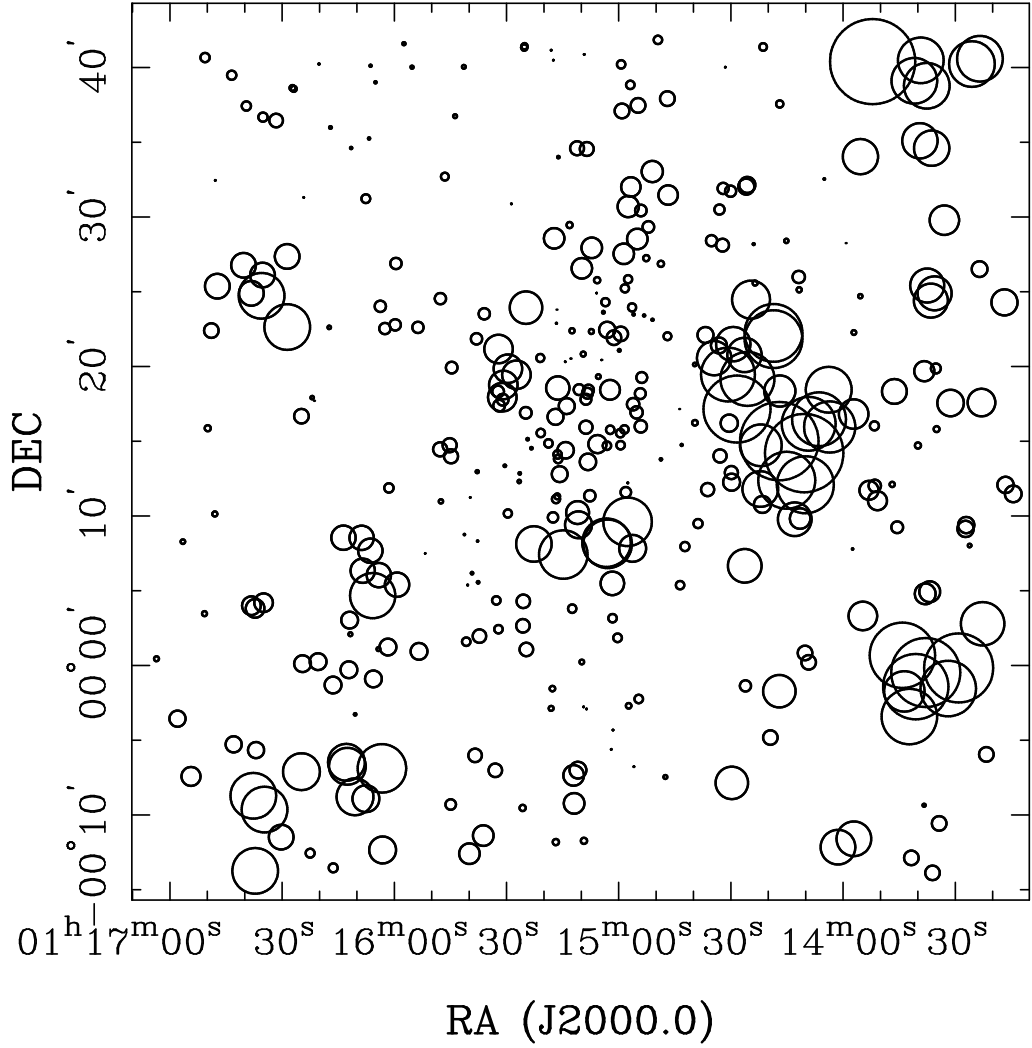}
\caption{The bubble plot shows the deviation
of the local velocity distribution for groups of 10 nearest neighbors
from  the overall velocity distribution. By involving $10^5$ simulations
the significance of existence of substructures reaches 
$P(\kappa_{10}>\kappa_{10}^{\rm obs})=0.000\%$.
\label{bub}}
\end{figure}

\clearpage

\begin{figure}
\epsscale{0.8}
\plotone{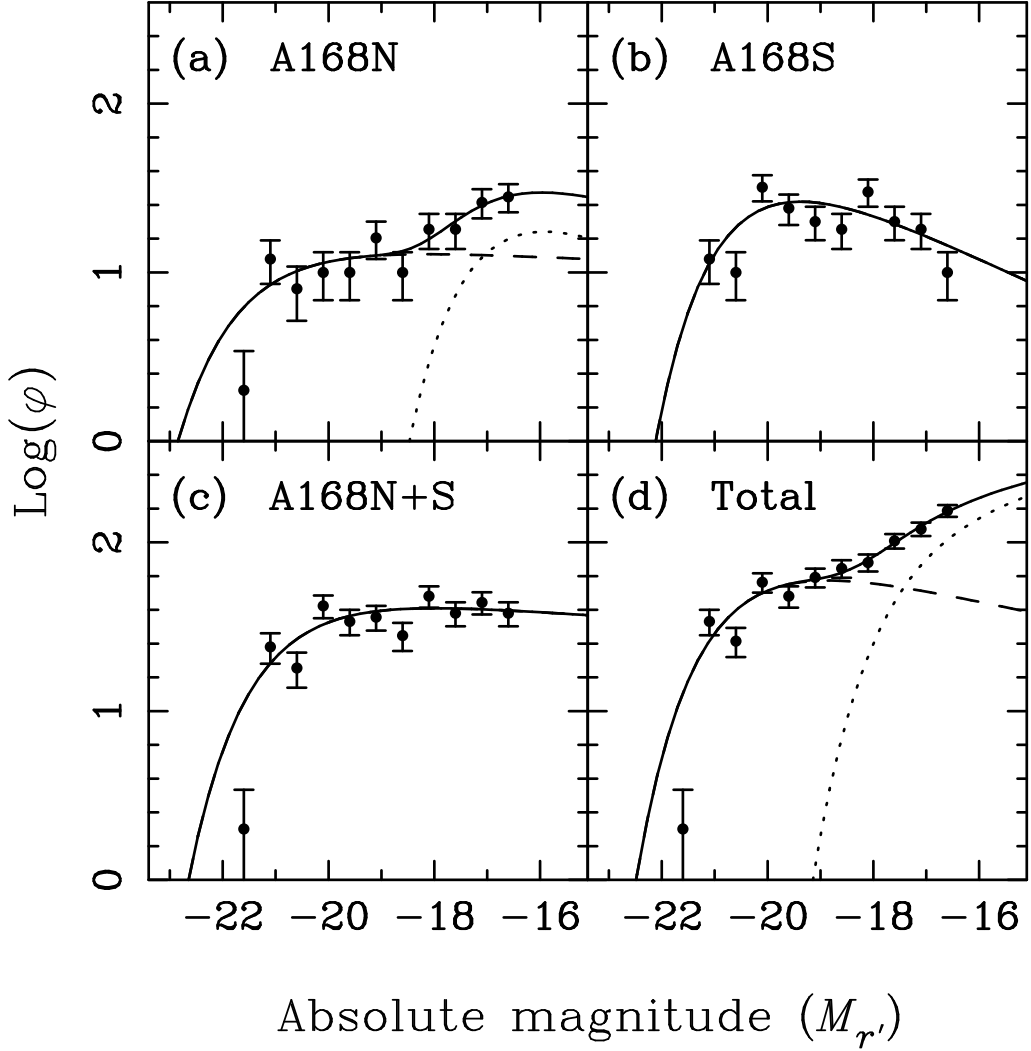}
\caption{The observed luminosity functions are fitted by DSF.  The
dashed lines and the dotted lines indicate the components of DSF and
the solid lines are the sum of them. Panel (a) and (b) are of the
north subcluster and south subcluster, respectively. The LF of the central
region is plotted in (c). In panel (d) we show the LF of all 376
members in the  whole field. For A168S and A168N+S (equal to the central 
region of \a168), the single Schechter
function gives a better representation than DSF.  The detailed
discussion on LFs is made in \S\ref{secLF}\label{figlum}. }
\end{figure}

\end{document}